\documentclass[conference]{IEEEtran}
\IEEEoverridecommandlockouts
\usepackage[noadjust]{cite}
\usepackage{amsmath}
\usepackage{array}
\usepackage{mdwmath}
\usepackage{amssymb}
\usepackage{mathtools}
\usepackage{algorithm}
\usepackage[noend]{algpseudocode}
\usepackage{eqparbox}
\usepackage{stfloats}
\usepackage{url}
\usepackage[pdftex]{graphicx}
\graphicspath{{../pdf/}{../jpeg/}}
\DeclareGraphicsExtensions{.pdf,.jpeg,.png}
\usepackage{epstopdf}
\usepackage{amsfonts}

\usepackage{color}

\usepackage{tikz}
\usepackage{pgfplots}
\usepackage{subcaption}
\usetikzlibrary{spy}

\newtheorem{lemma}{Lemma}
\newtheorem{theorem}{Theorem}
\newtheorem{remark}{Remark}
\newtheorem{corollary}{Corollary}

\begin{document}

\title{Performance Analysis of Zero-Forcing Precoding in  Multi-Cell One-Bit Massive MIMO Downlink}

\vspace{-.1in}
\author{\IEEEauthorblockN{Qurrat-Ul-Ain Nadeem and Anas Chaaban}
\IEEEauthorblockA{School of Engineering, The University of British Columbia, Kelowna, Canada}
Email: \{qurrat.nadeem, anas.chaaban\}@ubc.ca
}

\maketitle


\vspace{-.4in}
\begin{abstract}

This work investigates the downlink performance of a multi-cell massive multiple-input multiple-output (MIMO) system that employs one-bit analog-to-digital converters (ADCs) and digital-to-analog converters (DACs) in the receiving and transmitting radio frequency (RF) chains at each base station (BS) in order to reduce the power consumption. We utilize Bussgang decomposition to  derive the minimum mean squared error (MMSE) channel estimates at each BS based on the quantized received uplink training signals, and the asymptotic closed-form expressions of the   achievable downlink rates under one-bit quantized zero-forcing (ZF) precoding implemented using the estimated channels. The derived expressions explicitly show the impact of quantization noise, thermal noise, pilot contamination, and interference, and are utilized to study  the number of additional antennas needed at each BS of the one-bit MIMO system to perform as well as the conventional MIMO system.  Numerical results verify our analysis, and reveal that despite needing more antennas to achieve the same sum average rate, the one-bit massive MIMO system is more energy-efficient than the conventional system, especially at high sampling frequencies. 

\end{abstract}

\vspace{-.15in}
\section{Introduction}
\label{Sec:Intro}
\vspace{-.05in}

While the benefits of massive MIMO  scale with the number of antennas $M$ deployed at the base station (BS) \cite{massiveMIMOO}, the  power consumption and hardware cost associated with  active components, like power amplifiers, analogue-to-digital converters (ADCs) and  digital-to-analogue converters (DACs), that constitute the radio frequency (RF) chain connected to each antenna, also scale with $M$.  Moreover the power consumption of ADCs and DACs   increases exponentially with their resolution (in bits) and linearly with the sampling frequency \cite{ADC1, lit12}, with commercially available converters having $12$ to $16$ bits resolution consuming on the order of several watts \cite{lit9}.    The resolution of each ADC and DAC should therefore be limited  to keep the power consumption at the massive MIMO BSs within tolerable levels. 


 Motivated by this discussion, we will consider the simplest possible scenario of a  one-bit massive MIMO cellular network with BSs  employing one-bit ADCs and DACs, that consist of a simple comparator and consume very low power \cite{lit9}, and characterize the downlink achievable  rates  under zero-forcing (ZF) precoding implemented using imperfect channel state information (CSI). While  linear precoding schemes like  maximum ratio transmission (MRT)  and ZF have been shown to yield competitive performance  to the optimal high-complexity dirty paper coding scheme in conventional massive MIMO downlink where BSs employ full-resolution (FR) ADCs and DACs \cite{addref}, very little has been reported on the impact of low-resolution ADCs and DACs on their performance. 
 
 In this context,  the works in \cite{lit10} and \cite{lit11}  studied the achievable rates,  considering MRT precoding  and imperfect  CSI, in a single-cell one-bit massive MIMO  and a cell-free one-bit massive MIMO system respectively.  The authors in \cite{lit12}   derived asymptotic analytical  expressions of the signal-to-quantization-plus-interference-plus-noise ratio (SQINR) and  symbol error rate under one-bit quantized ZF precoding assuming perfect CSI and a single cell. Very recently, the authors in \cite{addref3}  considered a full-duplex massive MIMO cellular network with low-resolution ADCs and DACs at each BS, and derived spectral efficiency expressions under  MRT precoding. To the best of our knowledge, the downlink performance of one-bit quantized ZF precoding under imperfect CSI  has not been analyzed before in a single- or multi-cell setting.

In this work, we investigate the   downlink sum average rate performance of a multi-cell one-bit massive MIMO system under ZF precoding and imperfect CSI. The analysis is based on Bussgang decomposition \cite{Bussgang} that reformulates the nonlinear quantizer operation as a statistically equivalent linear system. We first derive the minimum mean squared error (MMSE) channel estimates at each BS  based on the received uplink training signals quantized by one-bit ADCs. Next the estimated CSI is used to implement ZF  and generate the transmit signals, which are  quantized by one-bit DACs. For this setting, we derive asymptotic closed-form expressions of the ergodic achievable downlink rates and study the extent of performance deterioration introduced by one-bit quantization. The derived expressions are used to study the ratio of the number of antennas at each BS in the one-bit cellular system to that at each BS in the conventional cellular system, required for both systems  to achieve the same sum average rate. The ratio turns out to be  $2.5 $ at low signal-to-noise ratio (SNR) values, while it is seen to decrease to one for any given SNR as the number of antennas grows large. Further, the numerical results  reveal that despite needing more antennas to achieve the same sum  rate, the one-bit  system is more energy efficient than the conventional system at high sampling frequencies.

%

\section{System Model}
\label{Sec:Sys}

\begin{figure}[t!]
\centering
\includegraphics[scale=.28]{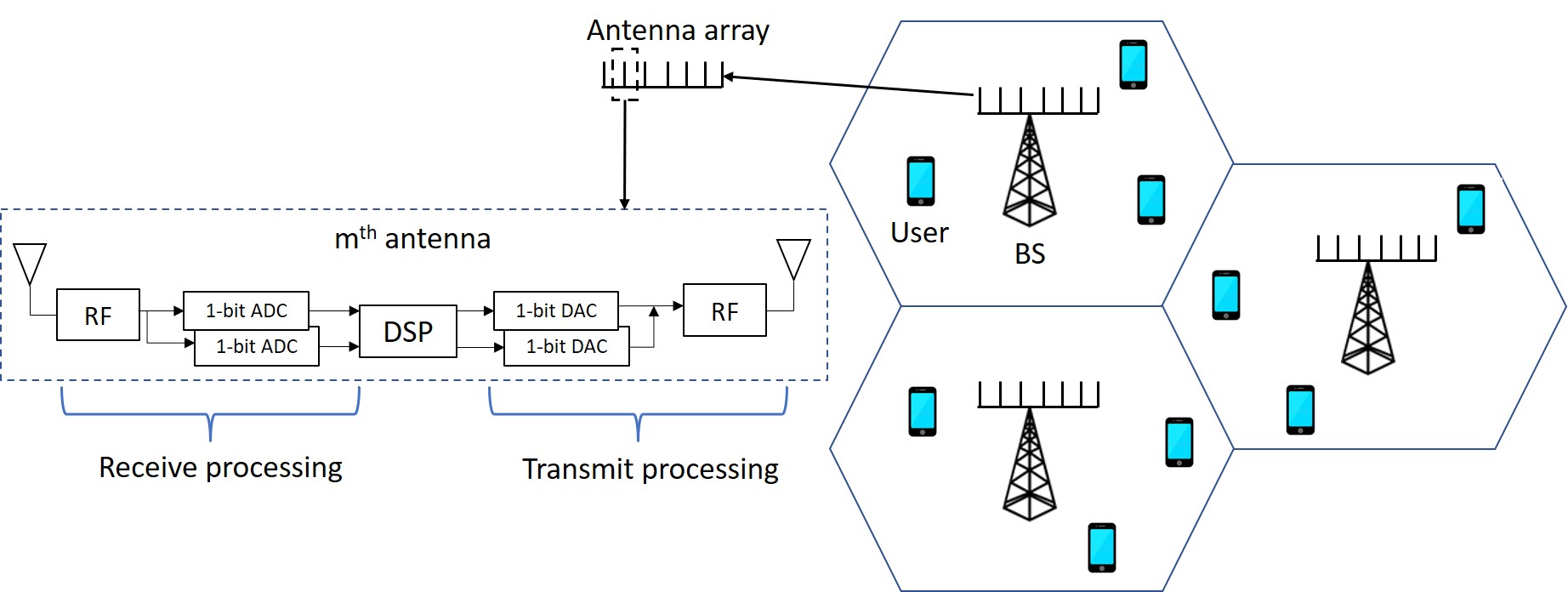}
\caption{Multi-cell one-bit massive MIMO system model. }
\label{model}
\end{figure}

We consider a multi-cell  massive MIMO system consisting of $ L > 1$ cells, with one $M$-antenna BS and $K\leq M$ single-antenna users in each cell.  BS $j$ wants to send information at rate $R_{jk}$ to user $k$ in  cell $j$. To do this, it constructs codewords with symbols $s_{jk} \in \mathbb{C}$ and combines them in a transmit signal vector $\mathbf{x}_j\in \mathbb{C}^{M\times 1}$ given as $\mathbf{x}_j=\mathbf{W}_j\mathbf{s}_j$, where $\mathbf{W}_j \in \mathbb{C}^{M \times K}$ is the linear precoder that BS $j$ applies to the vector of data symbols $\mathbf{s}_j=[s_{j1},\dots, s_{jK}]^T\in \mathbb{C}^{K\times 1}$, with the latter satisfying $\mathbb{E}[\mathbf{s}_j\mathbf{s}_j^H]=\mathbf{I}_K$. The RF chain associated with each antenna at each BS is equipped with a pair of one-bit ADCs and DACs as shown in Fig. \ref{model}, one for each of the  real and imaginary parts of the  signal. The real and imaginary parts of the  transmit signal $\mathbf{x}_j$ are  therefore converted into one bit representation (element-wise) based on the sign of each component, and then converted to analog using   one-bit DACs. The final transmit signal from BS $j$ is then written as \vspace{-.02in}
\begin{align}
\label{Tx}
&\tilde{\mathbf{x}}_j=Q(\mathbf{x}_j)=Q(\mathbf{W}_j\mathbf{s}_j),
\end{align}
where $Q(.)$ is the one-bit quantization operation defined as \vspace{-.02in}
\begin{align}
\label{Q}
&Q(\mathbf{a}) = \frac{1}{\sqrt{2}}\text{sign}(\Re(\mathbf{a})) + j \frac{1}{\sqrt{2}} \text{sign}(\Im(\mathbf{a})),
\end{align}
where $\Re(\mathbf{a})$ and  $\Im(\mathbf{a})$ represent the real and imaginary parts of $\mathbf{a}$, and $\text{sign}(\cdot)$ is the sign of their arguments. The elements of  $\tilde{\mathbf{x}}_j$ will  belong to  $\mathcal{R}=\frac{1}{\sqrt{2}}\{1+j, 1-j, -1+j, -1-j  \}$. 

The received signal  at all users in cell $j$ will be given as\vspace{-.05in}
\begin{align}
\label{Rx1}
&\mathbf{y}_j= \sum_{l=1}^L \sqrt{\eta_l} \mathbf{H}_{lj}^H \tilde{\mathbf{x}}_l+\mathbf{n}_j,
\end{align}
where $\mathbf{H}_{lj}=[\mathbf{h}_{lj1}, \dots, \mathbf{h}_{ljK}] \in \mathbb{C}^{M\times K}$, $\mathbf{h}_{ljk} \in \mathbb{C}^{M\times 1}$ is the channel from BS $l$ to user $k$ in cell $j$, $\mathbf{n}_j=[n_{j1}, \dots, n_{jK}]^T\in \mathbb{C}^{K \times 1}$, and $n_{jk} \sim \mathcal{CN}(0,\sigma^2)$ is the received noise at user $k$ in cell $j$.  Moreover $\eta_l$ is a normalization constant chosen to satisfy the average transmit power constraint at BS $l$ as $\mathbb{E}[||\sqrt{\eta_l}\tilde{\mathbf{x}}_l||^2]=P_t$. Since $\mathbb{E}[||\tilde{\mathbf{x}}_l||^2]=M$ due to \eqref{Q}, we  obtain $\eta_l=\frac{P_t}{M}$.  The channel matrix $\mathbf{H}_{lj}$ is modeled as  
\begin{align}
\label{channel}
&\mathbf{H}_{lj}=\mathbf{G}_{lj}\mathbf{D}_{lj}^{1/2},
\end{align}
 where $\mathbf{G}_{lj}=[\mathbf{g}_{lj1}, \dots, \mathbf{g}_{ljK}]\in \mathbb{C}^{M\times K}$ captures the small-scale fading,  and $\mathbf{D}_{lj}=\text{diag}(\beta_{lj1}\dots, \beta_{ljK}) \in \mathbb{C}^{K\times K}$ captures the large-scale fading. The entries of $\mathbf{g}_{ljk}$ are independently and identically distributed (i.i.d.) complex Gaussian random variables, with zero mean and unit variance. The coefficients  $\beta_{ljk}$ represent the channel attenuation factors. 
 
 In the next section,  we will outline the channel estimation done at BS $j$ to obtain an estimate $\hat{\mathbf{H}}_{jj}$ of the channel matrix $\mathbf{H}_{jj}=[\mathbf{h}_{jj1},\dots, \mathbf{h}_{jjK}]$ to the users in its cell. This CSI is needed by the BS  to implement  precoding and construct the quantized transmit signal in \eqref{Tx}. We consider  ZF precoding in this work, which is well-known for its interference suppression capability, and is implemented using the estimated channels from the next section,  as $\mathbf{W}_j=\hat{\mathbf{H}}_{jj}(\hat{\mathbf{H}}_{jj}^H \hat{\mathbf{H}}_{jj})^{-1}$.

\section{Uplink Channel Estimation}
\label{Sec:Sys2}

BS $j$ obtains an estimate of $\mathbf{H}_{jj}=[\mathbf{h}_{jj1},\dots, \mathbf{h}_{jjK}] \in \mathbb{C}^{M\times K}$ in an uplink training phase of length $\tau_p$ symbols at the start of each coherence block, in which the $K$ users in  cell $j$  transmit mutually orthogonal pilot sequences, represented as $\boldsymbol{\Phi}_j=[\boldsymbol{\phi}_{j1},\dots, \boldsymbol{\phi}_{jK}] \in \mathbb{C}^{\tau_p \times K}$,  satisfying  $\boldsymbol{\Phi}_j^H\boldsymbol{\Phi}_j=\mathbf{I}_K$. The same set of  pilot sequences is transmitted by the $K$ users in every cell resulting in the channel estimate to be corrupted by pilot contamination. The received training signal $\mathbf{Y}_j^p\in \mathbb{C}^{M \times \tau_p}$ at BS $j$ is  given as $\mathbf{Y}_j^p=\sum_{l=1}^L\sqrt{\rho_p \tau_p} \mathbf{H}_{jl} \boldsymbol{\Phi}_l^T + \mathbf{N}^p_{j}$,  where $\rho_p$ is the uplink SNR  and $\mathbf{N}_j^p \in \mathbb{C}^{M \times \tau_p}$ has i.i.d. $\mathcal{CN}(\mathbf{0}, \mathbf{I}_M)$ columns representing the  noise. Next we write $\mathbf{y}_j^p=\text{vec}(\mathbf{Y}_j^P)$ as \vspace{-.15in}
\begin{align}
\label{URx1}
&\mathbf{y}_j^p=\sum_{l=1}^L \bar{\boldsymbol{\Phi}}_l \mathbf{h}_{jl} +\mathbf{n}_j^p,
\end{align}
where $\bar{\boldsymbol{\Phi}}_l=(\boldsymbol{\Phi}_l \otimes \sqrt{\rho_p \tau_p} \mathbf{I}_M)\in \mathbb{C}^{M\tau_p \times MK}$,  $ \mathbf{h}_{jl}=\text{vec}(\mathbf{H}_{jl})$, $\mathbf{n}_j^p=\text{vec}(\mathbf{N}_{j}^p)$, and $\otimes$ represents the Kronecker product.

The RF chain with each antenna at the BS is equipped with a pair of one-bit ADCs as shown in Fig. \ref{model}, that separately quantize the real and imaginary parts of the received  signal  to one-bit representation based on their sign. The quantized received training signal after one-bit ADCs is thus given as \vspace{-.05in}
\begin{align}
\label{URx2}
&\mathbf{r}_j^p =Q(\mathbf{y}_j^p)=Q\left(\sum_{l=1}^L  \bar{\boldsymbol{\Phi}}_l \mathbf{h}_{jl} +\mathbf{n}_j^p\right),
\end{align}
where $Q(.)$ is defined in \eqref{Q}, and $\mathbf{r}_j^p$ takes values from $\mathcal{R}$.

\subsection{Bussgang Decomposition}


Quantizing the training signal introduces a distortion $Q(\mathbf{y}_j^p)-\mathbf{r}_j^p$ that is correlated with the input $\mathbf{y}_j^p$ to the ADCs. However, for Gaussian inputs, Bussgang’s theorem \cite{Bussgang} allows us to decompose the quantized signal into a linear function of the input to the quantizer and a distortion term that is uncorrelated with the input \cite{addref, lit9}. The resulting linear representation of the non-linear quantization operation is statistically equivalent up to the second moments of the data and therefore facilitates  ergodic rate analysis.  Specifically, the Bussgang decomposition of $\mathbf{r}_j^p $  in \eqref{URx2} is given as \cite{lit9, lit10}\vspace{-.05in}
\begin{align}
\label{URx3}
&\mathbf{r}_j^p =\mathbf{A}_j^p\mathbf{y}_j^p+\mathbf{q}_j^p ,
\end{align}
 where the matrix $\mathbf{A}_j^p$ is the linear operator chosen to satisfy $\mathbb{E}[\mathbf{y}_j^p \mathbf{q}_j^{p^H}]=\mathbf{0}$ as $\mathbf{A}_j^p= \mathbf{R}^H_{\mathbf{y}_j^p \mathbf{r}_j^p} \mathbf{R}^{-1}_{\mathbf{y}_j^p \mathbf{y}_j^p}$ \cite{lit9},  and $\mathbf{q}_j^{p}$ is the uncorrelated quantizer noise. Further, for one-bit quantization and Gaussian inputs,  $\mathbf{R}_{\mathbf{y}_j^p \mathbf{r}_j^p}=\sqrt{\frac{2}{\pi}} \mathbf{R}^H_{\mathbf{y}_j^p \mathbf{y}_j^p} \text{diag}(\mathbf{R}_{\mathbf{y}_j^p \mathbf{y}_j^p})^{-1/2}$ \cite{Bussgang},  \cite[Chap. 10]{Bussgang1}, which yields\vspace{-.05in}
 \begin{align}
 \label{A}
&\mathbf{A}_j^p= \sqrt{\frac{2}{\pi}}\text{diag}(\mathbf{R}_{\mathbf{y}_j^p \mathbf{y}_j^p})^{-1/2},
\end{align}
where $\mathbf{R}^H_{\mathbf{y}_j^p \mathbf{y}_j^p}$ is the auto-covariance matrix of  $\mathbf{y}_j^p$ in \eqref{URx1}, and $\text{diag}(\mathbf{C})$ denotes a diagonal square matrix with main-diagonal elements  equal to those of $\mathbf{C}$. It is also useful to provide here the covariance matrix of $\mathbf{q}_j^p$, that can be written for a one-bit quantizer using the arcsin law as \cite{lit9, lit11, lit12} \vspace{-.05in}
\begin{align}
\label{C_q}
&\mathbf{R}_{\mathbf{q}_j^p \mathbf{q}_j^p}=\frac{2}{\pi} (\arcsin(\mathbf{B})+j \arcsin(\mathbf{C})  )-\frac{2}{\pi}(\mathbf{B}+j\mathbf{C}),
\end{align}
where $\mathbf{B}=\text{diag}(\mathbf{R}_{\mathbf{y}_j^p \mathbf{y}_j^p})^{-1/2} \Re(\mathbf{R}_{\mathbf{y}_j^p \mathbf{y}_j^p})  \text{diag}(\mathbf{R}_{\mathbf{y}_j^p \mathbf{y}_j^p})^{-1/2}$ and $\mathbf{C}=\text{diag}(\mathbf{R}_{\mathbf{y}_j^p \mathbf{y}_j^p})^{-1/2} \Im(\mathbf{R}_{\mathbf{y}_j^p \mathbf{y}_j^p})  \text{diag}(\mathbf{R}_{\mathbf{y}_j^p \mathbf{y}_j^p})^{-1/2}$.

Next we utilize these results to complete the Bussgang decomposition of the quantized training signal in \eqref{URx2}. We substitute \eqref{URx1} in \eqref{URx3} to write the Bussgang decomposition as \vspace{-.1in}
\begin{align}
\label{URx4}
&\mathbf{r}_j^p=\sum_{l=1}^L\mathbf{A}_j^p \bar{\boldsymbol{\Phi}}_l   \mathbf{h}_{jl} +\mathbf{A}_j^p\mathbf{n}_j^p+\mathbf{q}_j^p,
\end{align}
where to find $\mathbf{A}_j^p$  using \eqref{A}, we compute  $\mathbf{R}_{\mathbf{y}_j^p \mathbf{y}_j^p}$ as $\mathbf{R}_{\mathbf{y}_j^p \mathbf{y}_j^p}=\mathbb{E}\left[\mathbf{y}_j^p\mathbf{y}_j^{p^H}\right]=\sum_{l=1}^L\bar{\boldsymbol{\Phi}}_l \bar{\mathbf{D}}_{jl} \bar{\boldsymbol{\Phi}}_l^H+\mathbf{I}_{M\tau_p}$, where  $\bar{\mathbf{D}}_{jl}=\mathbf{D}_{jl}\otimes \mathbf{I}_M\in \mathbb{C}^{MK \times MK}$ and $\mathbf{D}_{jl}$ is defined in \eqref{channel}. The expression of $\mathbf{R}_{\mathbf{y}_j^p \mathbf{y}_j^p}$  indicates that the choice of $\boldsymbol{\Phi}_l$'s will affect the linear operator $\mathbf{A}_j^p$ as well as the quantization noise. In order to obtain analytically tractable expressions for $\mathbf{A}_j^p$ and the channel estimates, we consider $\tau_p=K$ and choose the $K$-dimensional identity matrix as each pilot matrix  as done in  \cite{lit10, lit11}. Note that investigating the impact of different choices of $\tau_p$ and $\boldsymbol{\Phi}_l$ on the quality of channel estimates under one-bit ADCs is an interesting research direction  \cite{lit9}, but is beyond the scope of this work. Using $\boldsymbol{\Phi}_l= \mathbf{I}_K$, we  obtain $\mathbf{R}_{\mathbf{y}_j^p \mathbf{y}_j^p}=\sum_{l=1}^L K\rho_p \bar{\mathbf{D}}_{jl}+\mathbf{I}_{MK}$, and compute $\mathbf{A}_j^p$  as
\begin{align}
\label{A_fin}
&\mathbf{A}_j^p=\bar{\mathbf{A}}_j^p\otimes \mathbf{I}_M,
\end{align}
where $\bar{\mathbf{A}}_{j}^p$ is a diagonal matrix with entries $[\bar{\mathbf{A}}_{j}^p]_{kk}=\bar{a}_{jk}=\sqrt{\frac{2}{\pi(\sum_{l=1}^L K\rho_p \beta_{jlk}+1)}}$. Further using $\mathbf{R}_{\mathbf{y}_j^p \mathbf{y}_j^p}$ in \eqref{C_q}, we have \vspace{-.1in}
\begin{align}
\label{C_qp}
\mathbf{R}_{\mathbf{q}_j^p\mathbf{q}_j^p}=\left(1-\frac{2}{\pi}  \right) \mathbf{I}_{MK}.
\end{align}
This completes the Bussgang decomposition of the quantized training signal $ \mathbf{r}_{j}^p$, with  \eqref{URx4} being the statistically equivalent linear representation of   \eqref{URx2} under the definition of $\mathbf{A}_j^p$ in \eqref{A_fin}.

\subsection{MMSE Estimation}
\vspace{-.02in}

The MMSE estimate of the channel vector $\mathbf{h}_{jj}$ at BS $j$  based on the quantized  training signal $\mathbf{r}_j^p$ is presented next.


\begin{lemma} \label{L1}  BS $j$  estimates $\mathbf{h}_{jj}\hspace{-.04in}=\hspace{-.02in}[\mathbf{h}_{jj1}^T,  \dots, \mathbf{h}_{jjK}^T]^T\hspace{-.05in}\in\hspace{-.02in} \mathbb{C}^{MK\times 1}$  using  the quantized  training signal in \eqref{URx4}  as \vspace{-.04in}
\begin{align}
\label{est1}
&\hat{\mathbf{h}}_{jj}=\sqrt{\rho_p K} \bar{\mathbf{D}}_{jj} {\mathbf{A}}_j^{p^H}  \mathbf{r}_j^p
\end{align}
where $\bar{\mathbf{D}}_{jj}=\mathbf{D}_{jj}\otimes \mathbf{I}_M$, and $\mathbf{A}_j^p$ is defined in \eqref{A_fin}.
\end{lemma}
\begin{IEEEproof}
The proof follows by applying the standard definition of the MMSE estimate \cite[equation (14)]{lit11}.
\end{IEEEproof}

 Although the channel estimate in \eqref{est1} is not Gaussian in general due to the quantization noise $\mathbf{q}_j^p$ that appears in $\mathbf{r}_j^p$, we can approximate it as Gaussian using  the Cramer's central limit theorem assuming $M$ is sufficiently large \cite{CLT, lit10, lit9}. Thus we consider the channel estimate to be distributed as $\hat{\mathbf{h}}_{jj}\sim \mathcal{CN}(\mathbf{0},\mathbf{R}_{\hat{\mathbf{h}}_{jj}\hat{\mathbf{h}}_{jj}} )$, where the covariance matrix of the estimate $\mathbf{R}_{\hat{\mathbf{h}}_{jj}\hat{\mathbf{h}}_{jj}}$ is given as $\mathbf{R}_{\hat{\mathbf{h}}_{jj}\hat{\mathbf{h}}_{jj}} =\mathbf{T}_{jj} \otimes  \mathbf{I}_{M}$, where $\mathbf{T}_{jj}$ is a diagonal matrix with entries\vspace{-.09in}
\begin{align}
\label{T}
[\mathbf{T}_{jj}]_{k,k}=t_{jjk}=\frac{2 \beta_{jjk}^2 \rho_p K}{\pi (\sum_{l=1}^L K\rho_p \beta_{jlk}+1)}.
\end{align}
Under  orthogonality property of  MMSE estimate, the channel estimate and the  estimation error defined as $\tilde{\mathbf{h}}_{jj}={\mathbf{h}}_{jj}-\hat{\mathbf{h}}_{jj}$, are uncorrelated, with  $\tilde{\mathbf{h}}_{jj}\sim \mathcal{CN}(\mathbf{0},\bar{\mathbf{D}}_{jj}-\mathbf{R}_{\hat{\mathbf{h}}_{jj}\hat{\mathbf{h}}_{jj}} )$.

To facilitate the analysis, we can extract the estimate of the channel from BS $j$ to user $k$ in cell $j$ from \eqref{est1} as $\hat{\mathbf{h}}_{jjk}= \sqrt{\rho_p K} \beta_{jjk} \bar{a}_{jk} \mathbf{r}_{jk}^p$, where $\mathbf{r}_{jk}^p=\sum_{l=1}^L  \sqrt{\rho_p K} \bar{a}_{jk} \mathbf{h}_{jlk}+\bar{a}_{jk}\mathbf{n}_{jk}^p+\mathbf{q}_{jk}^p$, and $\mathbf{n}_{jk}^p$ and $\mathbf{q}_{jk}^p$ are vectors of $(k-1)M+1$ to $kM$ entries of $\mathbf{n}_{j}^p$ and $\mathbf{q}_{j}^p$ respectively. It then follows that $\hat{\mathbf{h}}_{jjk}\sim \mathcal{CN}(\mathbf{0},t_{jjk}\mathbf{I}_M)$ where $t_{jjk}$ is defined in \eqref{T}, and  $\tilde{\mathbf{h}}_{jjk} \sim \mathcal{CN}(\mathbf{0},\tilde{t}_{jjk}\mathbf{I}_M)$, where $\tilde{t}_{jjk}=\beta_{jjk}-t_{jjk}$.

\begin{corollary}
\label{Cor1}
When BSs have full-resolution (FR) ADCs, the estimate of the channel from BS $j$ to user $k$ in cell $j$ is \vspace{-.05in}
\begin{align}
&\hat{\mathbf{h}}^{\rm FR}_{jjk}= \sqrt{\rho_p K} \beta_{jjk} \mathbf{r}^{p}_{jk}
\end{align}
 where $\mathbf{r}_{jk}^{p}=\sum_{l=1}^L  \sqrt{\rho_p K}  \mathbf{h}_{jlk}+\mathbf{n}_{jk}^p$.  It follows that $\hat{\mathbf{h}}^{\rm FR}_{jjk}\sim \mathcal{CN}(\mathbf{0},t^{\rm FR}_{jjk}\mathbf{I}_M)$ where $t^{\rm FR}_{jjk}=\frac{\beta_{jjk}^2 \rho_p K}{\sum_{l=1}^L K\rho_p \beta_{jlk}+1}$, and $\tilde{\mathbf{h}}^{\rm FR}_{jjk} \sim \mathcal{CN}(\mathbf{0},\tilde{t}_{jjk}^{\rm FR}\mathbf{I}_M)$ where $\tilde{t}^{\rm FR}_{jjk}=\beta_{jjk}-t^{\rm FR}_{jjk}$.
\end{corollary}
\begin{IEEEproof}
The corollary follows from \cite[Sec. II-C]{massiveMIMOO}.
\end{IEEEproof}

It is straightforward to see that $t_{jjk}=\frac{2}{\pi}t^{\rm FR}_{jjk}$, and therefore $\tilde{t}_{jjk} > \tilde{t}_{jjk}^{\rm FR}$.  The use of one-bit ADCs in the RF chains at the BSs  therefore deteriorates the accuracy of channel estimation, which will  decrease the system performance. 

\section{Downlink Achievable Rate Analysis}
\label{Sec:Asym}

In this section, we analyze the  ergodic achievable downlink rates under one-bit quantized ZF precoding.

\subsection{Bussgang Decomposition of Transmit Signal}

%

We again utilize Bussgang decomposition  to obtain a linear representation of the quantized transmit signal in \eqref{Tx}.  Even though the entries of $\mathbf{x}_j=\mathbf{W}_j\mathbf{s}_j$, which is the input to the quantizer in \eqref{Tx}, are not necessarily Gaussian, each element  of $\mathbf{x}_j$ is formed as a result of the linear mixture of  $K$ i.i.d. elements of the vector $\mathbf{s}_j$ and  can be approximated as Gaussian  using Cramer's central limit theorem \cite{CLT}  for large  $K$ \cite{lit10, lit12}. We  therefore apply  Bussgang theorem  to decompose the quantized  signal in   \eqref{Tx} into a linear function of the input $ \mathbf{x}_j$ to the quantizer and a quantization noise term $\mathbf{q}_j$ that is uncorrelated with  input as  \cite{lit10, lit12}, \cite[Theorem 2]{addref} \vspace{-.05in}
\begin{align}
\label{Tx2}
&\tilde{\mathbf{x}}_j=Q(\mathbf{x}_j)=\mathbf{A}_j \mathbf{x}_j+\mathbf{q}_j,
\end{align}
where $\mathbf{A}_j= \sqrt{\frac{2}{\pi}}\text{diag}\left( \mathbf{R}_{\mathbf{x}_j \mathbf{x}_j} \right)^{-1/2}$, and $\mathbf{R}_{\mathbf{x}_j \mathbf{x}_j}=\mathbb{E}_{\mathbf{s}_j}[\mathbf{x}_j \mathbf{x}_j^H]=\mathbb{E}_{\mathbf{s}_j}[\mathbf{W}_{j}\mathbf{s}_j \mathbf{s}_j^H\mathbf{W}_{j}^H]=\mathbf{W}_j\mathbf{W}_j^H$  \cite[Corollary 3]{addref}. Moreover using the arcsin law, the autocovariance matrix of $\mathbf{q}_j$ can be obtained as $\mathbf{R}_{\mathbf{q}_j \mathbf{q}_j}=\frac{2}{\pi} (\arcsin(\bar{\mathbf{B}})+j \arcsin(\bar{\mathbf{C}})  )-\frac{2}{\pi}(\bar{\mathbf{B}}+j\bar{\mathbf{C}})$, where $\bar{\mathbf{B}}=\text{diag}( \mathbf{R}_{\mathbf{x}_j \mathbf{x}_j} )^{-1/2} \Re( \mathbf{R}_{\mathbf{x}_j \mathbf{x}_j} )  \text{diag}( \mathbf{R}_{\mathbf{x}_j \mathbf{x}_j} )^{-1/2}$ and $\bar{\mathbf{C}}=\text{diag}( \mathbf{R}_{\mathbf{x}_j \mathbf{x}_j} )^{-1/2} \Im( \mathbf{R}_{\mathbf{x}_j \mathbf{x}_j} )  \text{diag}( \mathbf{R}_{\mathbf{x}_j \mathbf{x}_j} )^{-1/2}$  \cite{lit11, addref}. Next we approximate $ \mathbf{R}_{\mathbf{x}_j \mathbf{x}_j}=\mathbf{W}_j\mathbf{W}_j^H $ as a deterministic quantity under ZF precoding and find $\mathbf{A}_j$ to complete the Bussgang decomposition in \eqref{Tx2}. 

\begin{lemma}\label{LemmaZF}
Under ZF precoding, the Bussgang decomposition of the quantized transmit signal in \eqref{Tx}  for large $(M,K)$ values such that the ratio $M/K=c < \infty$,  is given as\vspace{-.07in}
\begin{align}
\label{Tx2ZF}
&\tilde{\mathbf{x}}_j=\mathbf{A}_j \mathbf{W}_j \mathbf{s}_j+\mathbf{q}_j
\end{align}
where $\mathbf{A}_j= \sqrt{\frac{2 K (c-1)^2}{\pi \zeta_j}}\mathbf{I}_M$,  $\mathbf{R}_{\mathbf{q}_j \mathbf{q}_j}= \left(1-\frac{2}{\pi} \right) \mathbf{I}_M$, and $\zeta_j=\frac{1}{K}\sum_{k=1}^K \frac{1}{t_{jjk}}$, where $t_{jjk}$ is defined in Lemma \ref{L1}.
\end{lemma}
\begin{IEEEproof}
We utilize  \cite[(34)]{lit12}  to obtain an asymptotic approximation for  $ \mathbf{R}_{\mathbf{x}_j \mathbf{x}_j} =\mathbf{W}_j\mathbf{W}^H_j=\hat{\mathbf{H}}_{j}(\hat{\mathbf{H}}^H_{j}\hat{\mathbf{H}}_{j})^{-2}\hat{\mathbf{H}}_{j}^H$  under ZF precoding, which is very  tight for moderate system sizes as well, and use it to compute  $\mathbf{A}_j$ and $\mathbf{R}_{\mathbf{q}_j \mathbf{q}_j}$. 
\end{IEEEproof}

\vspace{-.02in}
\subsection{Achievable Rates}
\vspace{-.03in}


 We now outline the achievable  rates at the users and develop  closed-form expressions  for them under one-bit and conventional massive MIMO settings.  To this end, we  utilize the  decomposition of $\tilde{\mathbf{x}}_j$ in \eqref{Tx2ZF} to write the received signal at user $k$ in cell $j$ using \eqref{Rx1} as ${y}_{jk}=\sum_{l=1}^L \sqrt{\eta_l}\mathbf{h}_{ljk}^H \mathbf{A}_l \mathbf{W}_{l} \mathbf{s}_l+\sum_{l=1}^L \sqrt{\eta_l}\mathbf{h}_{ljk}^H \mathbf{q}_l+{n}_{jk}$. Since the users do not have  channel estimates, we provide an ergodic achievable  rate  based on the technique developed in \cite{AR}, that exploits the fact that the effective channel $\mathbf{h}_{jjk}^H \mathbf{A}_j \mathbf{w}_{jk}$ of user $k$ in cell $j$ approaches its average value $\mathbb{E}[\mathbf{h}_{jjk}^H \mathbf{A}_j \mathbf{w}_{jk} ]$ as $M$ grows large due to channel hardening. Hence, asymptotically it is sufficient for each user to only have   statistical CSI (i.e. knowledge of $\mathbb{E}[\mathbf{h}_{jjk}^H \mathbf{A}_j \mathbf{w}_{jk} ]$). The main idea then is to decompose  ${y}_{jk}$  as 

\footnotesize
\begin{align}
&{y}_{jk}=\hspace{-.03in}\underbrace{\sqrt{\eta_j}\mathbb{E}[\mathbf{h}_{jjk}^H \mathbf{A}_j \mathbf{w}_{jk} ] {s}_{jk}}_{\text{Desired signal}} \hspace{-.03in}+\hspace{-.03in}\underbrace{\sqrt{\eta_j}(\mathbf{h}_{jjk}^H \mathbf{A}_j \mathbf{w}_{jk}\hspace{-.03in} -\hspace{-.03in}\mathbb{E}[\mathbf{h}_{jjk}^H \mathbf{A}_j \mathbf{w}_{jk} ]) {s}_{jk}}_{\text{Channel gain uncertainty}}\nonumber  \\
\label{Rx5}
&+\underbrace{\sum_{(l,m)\neq (j,k)} \sqrt{\eta_l}\mathbf{h}_{ljk}^H \mathbf{A}_l \mathbf{w}_{lm} {s}_{lm}}_{\text{Inter-user interference}}+\underbrace{\sum_{l=1}^L \sqrt{\eta_l}\mathbf{h}_{ljk}^H \mathbf{q}_l}_{\text{Quantization noise}}+\underbrace{{n}_{jk}}_{\text{Thermal noise}} 
\end{align} \normalsize
and assume that the average effective channel $\mathbb{E}[\mathbf{h}_{jjk}^H \mathbf{A}_j \mathbf{w}_{jk} ]$ can be perfectly learned at  user $k$ in cell $j$. The sum of the last four terms in \eqref{Rx5} is  considered as effective additive noise. Treating this noise   as uncorrelated Gaussian as a worst-case,   user $k$ in cell $j$ can achieve the  ergodic  rate  \cite[Theorem 1]{AR}
\begin{align}
\label{rate}
&R_{jk}=\log_2(1+\gamma_{jk})
\end{align}
where $\gamma_{jk}$ is the associated SQINR obtained using  \eqref{Rx5} as \vspace{-.06in}
\begin{align}
\label{SQINR}
&\gamma_{jk}=\frac{\text{DS}_{jk}}{\text{CU}_{jk}+\text{QN}_{jk}+\text{IUI}_{jk}+\text{TN}_{jk}}
\end{align}
where  $\text{DS}_{jk}={\eta_j}|\mathbb{E}[\mathbf{h}_{jjk}^H \mathbf{A}_j \mathbf{w}_{jk} ]|^2$ is the power of the average desired signal,   $\text{CU}_{jk}=\eta_j\text{Var}[\mathbf{h}_{jjk}^H \mathbf{A}_j \mathbf{w}_{jk} ]$ is the average channel gain uncertainty power,  $\text{QN}_{jk}=\sum_{l=1}^L \eta_l\mathbb{E}[\mathbf{h}_{ljk}^H \mathbf{C}_{\mathbf{q}_l\mathbf{q}_l} \mathbf{h}_{ljk}]$  is the average quantization noise power,  $\text{IUI}_{jk}=\sum_{(l,m)\neq (j,k)} \eta_l \mathbb{E}[|\mathbf{h}_{ljk}^H \mathbf{A}_l \mathbf{w}_{lm}|^2]$ is the average inter-user interference power,  and $\text{TN}_{jk}=\sigma^2$ is the thermal noise power. The  sum average rate  is then given as \vspace{-.09in}
\begin{align}
\label{Rsum}
&R_{\rm sum}=\sum_{j=1}^L \sum_{k=1}^K R_{jk}.
\end{align}
Note that these definitions of the ergodic achievable downlink rate in \eqref{rate} and SQINR in \eqref{SQINR}  will be used for performance evaluation based on Monte-Carlo simulations in Sec. V.  To yield explicit theoretical insights into the impact of one-bit quantization on the sum average rate, we   derive the expectations in \eqref{SQINR}  in closed-form, resulting in an analytical expression for \eqref{rate} that is presented in the following theorem.

\begin{theorem}
\label{Th_ZF}
Consider a one-bit massive MIMO cellular network with BSs equipped with one-bit ADCs and DACs. Then under ZF precoding and large $(M,K)$ values such that $\frac{M}{K}=c$ is finite, the ergodic achievable  rate in \eqref{rate} and  SQINR in \eqref{SQINR} at user $k$ in cell $j$ are given in closed-form  as \vspace{-.05in}
\begin{align}
R^{\rm  one}_{jk}&=\log_2(1+\gamma^{\rm one}_{jk}), \\
\label{SQINR_ZF}
\gamma^{\rm  one}_{jk}&=\frac{1}{\overline{\text{CU}}^{\rm  one}_{jk}+\overline{\text{QN}}^{\rm  one}_{jk}+\overline{\text{IUI}}^{\rm  one}_{jk}+\overline{\text{PC}}^{\rm  one}_{jk}+\overline{\text{TN}}^{\rm  one}_{jk}},
\end{align}
where $\overline{\text{CU}}^{\rm  one}_{jk}=\frac{\beta_{jjk}-t_{jjk}}{(M-K) t_{jjk}}$ is the normalized average channel gain uncertainty power,   $\overline{\text{QN}}^{\rm  one}_{jk}=\sum_{l=1}^L \left(1-\frac{2}{\pi}  \right) \frac{\pi M   \beta_{ljk} \zeta_j}{2K (c-1)^2}$ is the  normalized average  quantization noise power,   $\overline{\text{IUI}}^{\rm  one}_{jk}=\sum_{m\neq k}^K\frac{(\beta_{jjk}-t_{jjk})}{  t_{jjm} (M-K)}+\sum_{l\neq j}^L\sum_{m\neq k}^K\frac{\zeta_j \beta_{ljk}}{\zeta_l  t_{llm} (M-K)}+\sum_{l\neq j}^L \frac{\zeta_j \beta_{ljk}}{\zeta_l  t_{llk} (M-K)}\left(1-\frac{t_{llk}\beta_{ljk}}{\beta_{llk}^2}  \right)$ is the normalized average inter-user interference power,   $\overline{\text{PC}}^{\rm  one}_{jk}=\sum_{l\neq j}^L \frac{\zeta_j \beta_{ljk}^2}{\zeta_l  \beta_{llk}^2}$ is the normalized average pilot contamination power, and   $\overline{\text{TN}}^{\rm  one}_{jk}=\frac{\pi M \sigma^2 \zeta_j}{2K P_t (c-1)^2}$ is the normalized average thermal noise power (all normalized by the power of the average desired signal).
\end{theorem}
\begin{IEEEproof}
The proof follows by using $\mathbf{A}_j=\sqrt{\frac{2 K (c-1)^2}{\pi \zeta_{j}}} \mathbf{I}_M$,  $\eta_j=\frac{P_t}{M}$,  the channel in \eqref{channel}, the estimates in Lemma \ref{L1},  $\mathbf{H}_{jj}^H \mathbf{W}_j=\mathbf{I}_K+\tilde{\mathbf{H}}_{jj}^H \mathbf{W}_j$, and the observation that the estimates of the channels at BS $l$ to user $k$ in cell $l$ and to user $k$ in cell $j$ are correlated due to pilot contamination, to compute the terms in  \eqref{SQINR}. All terms in the denominator of the resulting expression are divided by the expression of $\text{DS}_{jk}$, and represented as $\overline{\text{CU}}_{jk}$, $\overline{\text{QN}}^{\rm }_{jk}$, $\overline{\text{IUI}}^{\rm }_{jk}$, $\overline{\text{PC}}^{\rm }_{jk}$, and $\overline{\text{TN}}^{\rm }_{jk}$. 
\end{IEEEproof}

Next  we present the closed-form expression of the  achievable rate  in \eqref{rate} for the conventional massive MIMO network.
\begin{corollary}
\label{Cor_ZFconv}
Consider the conventional massive MIMO cellular network employing FR ADCs and DACs. Then under ZF precoding, the ergodic  rate $R_{jk}$ in \eqref{rate} and the associated SINR $\gamma_{jk}$ of user $k$ in cell $j$ are given  in closed-form as
\begin{align}
R^{\rm conv}_{jk}&=\log_2(1+\gamma^{\rm  conv}_{jk}), \\
\label{SQINR_ZFconv}
\gamma^{\rm conv}_{jk}&=\frac{1}{\overline{\text{CU}}^{\rm conv}_{jk}+\overline{\text{IUI}}^{\rm conv}_{jk}+\overline{\text{PC}}^{\rm conv}_{jk}+\overline{\text{TN}}^{\rm conv}_{jk}}, \end{align}
where $\overline{\text{CU}}^{\rm conv}_{jk}=\frac{\beta_{jjk}-t^{\rm FR}_{jjk}}{(M-K) t^{\rm FR}_{jjk}}$, $\overline{\text{IUI}}^{\rm  conv}_{jk}=\sum_{m\neq k}^K\frac{(\beta_{jjk}-t^{\rm FR}_{jjk})}{  t^{\rm FR}_{jjm} (M-K)}+\sum_{l\neq j}^L\sum_{m\neq k}^K\frac{\zeta^{\rm FR}_j \beta_{ljk}}{\zeta^{\rm FR}_l  t^{\rm FR}_{llm} (M-K)}+ \sum_{l\neq j}^L \frac{\zeta^{\rm FR}_j \beta_{ljk}}{\zeta^{\rm FR}_l  t^{\rm FR}_{llk} (M-K)}\left(1-\frac{t^{\rm FR}_{llk}\beta_{ljk}}{\beta_{llk}^2}  \right)$, $\overline{\text{PC}}^{\rm  conv}_{jk}=\sum_{l\neq j}^L \frac{\zeta^{\rm FR}_j \beta_{ljk}^2}{\zeta^{\rm FR}_l  \beta_{llk}^2}$, and $\overline{\text{TN}}^{\rm  conv}_{jk}=\frac{ \sigma^2 K \zeta^{\rm FR}_j}{P_t (M-K)}$, with $t^{\rm FR}_{jjk}$ defined in Corollary \ref{Cor1}  and $\zeta^{\rm FR}_j=\frac{1}{K}\sum_{k=1}^K \frac{1}{t^{\rm FR}_{jjk}}$.
\end{corollary}

Comparing the results in Theorem \ref{Th_ZF} and Corollary \ref{Cor_ZFconv}, we see that using one-bit ADCs and DACs not only introduces a quantization noise term $\overline{\text{QN}}_{jk}$ in $\gamma_{jk}$, but it also increases the  noise and interference terms as $\frac{\overline{\text{TN}}^{\rm  one}_{jk}}{\overline{\text{TN}}^{\rm conv}_{jk}}=\frac{\pi^2 M}{4K (c-1)}\approx \frac{\pi^2}{4}$, $\frac{\overline{\text{CU}}^{\rm one}_{jk}}{\overline{\text{\rm CU}}^{\rm  conv}_{jk}}=\frac{\pi}{2} \left(\frac{\beta_{jjk}-\frac{2}{\pi}t_{jjk}^{\rm FR}}{\beta_{jjk}-t_{jjk}^{\rm FR}}  \right)> \frac{\pi}{2}$,    and $\frac{\overline{\text{IUI}}^{\rm  one}_{jk}}{\overline{\text{IUI}}^{\rm  conv}_{jk}}  =\frac{\pi}{2}\frac{(M-K)\overline{\text{IUI}}_{jk}^{\rm  conv} +\sum_{m\neq k}^K \left(1-\frac{2}{\pi}\right)\frac{t_{jjk}^{\rm FR}}{t_{jjm}^{\rm FR}} +\sum_{l \neq j} \left(1-\frac{2}{\pi}\right) \frac{\zeta_j^{\rm FR} \beta^2_{ljk}}{\zeta_l^{\rm FR} \beta_{llk}^2 }}{(M-K)\overline{\text{IUI}}_{jk}^{\rm  conv}}   > \frac{\pi}{2}$, resulting in reduced  rates under the setting of Theorem  \hspace{-.04in} \ref{Th_ZF}.  \hspace{-.05in} Interestingly, \hspace{-.05in} pilot  \hspace{-.05in} contamination   \hspace{-.04in} to  \hspace{-.04in} desired signal  \hspace{-.03in} energy ratio is unaffected by one-bit quantization as $\overline{\text{PC}}^{\rm one}_{jk} \hspace{-.02in}= \overline{\text{PC}}^{\rm conv}_{jk}$.



Finally we show that the performance under both settings considered in Theorem \ref{Th_ZF} and Corollary \ref{Cor_ZFconv} converges to the same limit as $M\rightarrow \infty$ while the other variables are fixed.
\begin{corollary}
\label{Cor_ZFlimit}
The ergodic achievable downlink rates for both settings above converge as $R_{jk}\xrightarrow[M\rightarrow \infty]{}R^{\rm  \infty}_{jk}$ , where  $R^{\rm \infty}_{jk}=\log_2\left(1+\frac{1}{\overline{\text{PC}}_{jk}}\right)$,  $\overline{\text{PC}}_{jk}=\sum_{l\neq j}^L \frac{\bar{\zeta}_j \beta_{ljk}^2}{\bar{\zeta}_l  \beta_{llk}^2}$ represents the  average pilot contamination power to average desired signal power ratio,  $\bar{\zeta}_{j}=\sum_{k=1}^K \frac{1}{c_{jjk}}$, and $c_{jjk}=\frac{\beta_{jjk}^2}{\sum_{l=1}^L K \rho_p \beta_{jlk}+1}$.
\end{corollary}

Therefore the effects of channel  uncertainty, quantization noise, thermal noise, and interference vanish as $M\rightarrow \infty$, while pilot contamination remains the only performance limitation under both settings. This  also implies that by using a larger number of antennas equipped with low power one-bit ADCs and DACs in the one-bit massive MIMO network, we can compensate for  quantization noise and approach the performance of   conventional massive MIMO  as studied next.

\subsection{How Many More Antennas  are Needed in One-Bit MIMO?}\vspace{-.03in}

%

We denote the number of antennas at each BS and the achievable sum average rate of the one-bit and conventional cellular systems as ($M^{\rm one}$, $R_{\rm sum}^{\rm one}$) and ($M^{\rm conv}$, $R_{\rm sum}^{\rm  conv}$) respectively. Our goal in this section is to study the  ratio $\kappa=\frac{M^{\rm one}}{M^{\rm conv}}$ required for the one-bit massive MIMO system to achieve the same sum average rate as the conventional massive MIMO system with $M^{\rm conv}$ antennas. In the low SNR regime, i.e. for small values of $\frac{P_t}{\sigma^2}$, we obtain  $\kappa$ explicitly as follows.


\begin{corollary}\label{Corspec1}
At low SNR values, the ratio $\kappa$ required for  one-bit  massive MIMO cellular system to achieve the sum average rate of  conventional massive MIMO cellular system with $M^{\rm conv}$ antennas at each BS is  $\kappa=\frac{M^{\rm  one}}{M^{\rm  conv}}\approx \frac{\pi^2}{4}\approx 2.5$.
\end{corollary}
\begin{IEEEproof}
The proof follows by simplifying  \eqref{SQINR_ZF}  and  \eqref{SQINR_ZFconv}  for small $\frac{P_t}{\sigma^2}$, and finding $\kappa$ to guarantee $R_{\rm sum}^{\rm one}=R_{\rm sum}^{\rm conv}$.
\end{IEEEproof}


While we can not get a closed-form expression for $\kappa$ at moderate to high SNR values, we will find it numerically in the simulations through a simple search over the interval $[1, \infty )$ to guarantee that $R_{\rm sum}^{\rm one}=R_{\rm sum}^{\rm conv}$, and will observe it to be  $>2.5$  for moderate values of $M^{\rm conv}$. This is because as the SNR increases, the quantization noise term comes into play  in $\gamma^{\rm one}_{jk}$, and the channel gain uncertainty and inter-user interference  terms also become dominant and are significantly increased under  one-bit implementation. This  results in an overall larger decrease in achievable sum average rate in the one-bit  MIMO setting, requiring a higher $\kappa$  to compensate for it and achieve $R_{\rm sum}^{\rm  conv}$. However as $M^{\rm conv}$ increases to larger numbers,  $\kappa$  decreases  and approaches one as outlined next.


\begin{remark}
As $M^{\rm conv}\rightarrow \infty$,  $\kappa=M^{\rm one}/M^{\rm  conv}\rightarrow 1$ since    $R_{\rm sum}^{\rm one}$ and $R_{\rm sum}^{\rm conv}$ both converge to $\sum_{l=1}^L \sum_{j=1}^K R^{\infty}_{jk}$ as  shown in Corollary \ref{Cor_ZFlimit}. Therefore the impact of one-bit quantization   becomes smaller as we work with larger antenna arrays.
\end{remark}


\section{Simulation Results}
\label{Sec:Sim}
We consider $L=4$ cells (unless otherwise stated in the figure) with  Cartesian coordinates of the BSs set as $(0,0,0)$, $(525, 0, 0)$, $(0, 525, 0)$, and $(525, 525, 0)$ (in metres). The BS in each cell has $M$ antennas serving $K$ users distributed uniformly on a circle  of radius $250$ metres around it \cite{massiveMIMOO}. Moreover  $\sigma^2=-80$\rm{dBm}, $\rho_p=\frac{1}{\sigma^2}$,  $\beta_{jlk}=\frac{10^{-3}}{d_{jlk}^\alpha}$,  $\alpha=3$, and $d_{jlk}$ is the distance between BS $j$ and user $k$ in cell $l$. 

We first validate the closed-form expressions of the achievable rates in Fig. \ref{Fig2}, where we plot the   sum ergodic rate per user given as $\frac{1}{KL}R_{\rm sum}$.  The theoretical (Th) results  are plotted  using the expressions of  $R_{jk}$ in  Theorem \ref{Th_ZF} and Corollary \ref{Cor_ZFconv} for the one-bit and conventional massive MIMO scenarios respectively. The Monte-Carlo (MC) simulated curves are plotted by computing  $R_{jk}$ in \eqref{rate}  for both scenarios. A perfect match between the MC simulated and theoretical results can be seen, even for moderate system dimensions.  As expected, there is a performance degradation when we use one-bit ADCs and DACs, with the decrease being more significant for a smaller number of cells. This is because the intra-cell interference becomes noticeable when compared to inter-cell interference for $L=2$, and is more effectively combated by conventional ZF  than one-bit quantized ZF precoding. 

\begin{figure*}[!t]
\begin{minipage}[b]{0.32\linewidth}
\centering
\tikzset{every picture/.style={scale=.95}, every node/.style={scale=.8}}
%
%
\definecolor{mycolor1}{rgb}{0.74902,0.00000,0.74902}%
\definecolor{mycolor2}{rgb}{0.60000,0.20000,0.00000}%
\definecolor{mycolor3}{rgb}{0.00000,0.49804,0.00000}%
\begin{tikzpicture}

\begin{axis}[%
width=.85\textwidth,
height=.7\textwidth,
scale only axis,
xmin=-30,
xmax=20,
xlabel style={font=\color{white!15!black}},
xlabel={Average transmit power $P_t$ (dB)},
ymin=0,
ymax=4.3,
ylabel style={font=\color{white!15!black}},
ylabel={Sum average rate per user (bps/Hz)},
axis background/.style={fill=white},
title style={font=\bfseries},
xmajorgrids,
ymajorgrids,
legend style={at={(axis cs: 20,0)},anchor=south east,legend cell align=left,align=left,draw=white!15!black, /tikz/column 2/.style={
                column sep=5pt,
            }},]

\addplot [color=blue, line width=1.0pt]
  table[row sep=crcr]{%
-30	0.035262200958826\\
-28	0.0551340187719707\\
-26	0.0855655102761823\\
-24	0.13132409556032\\
-22	0.198319209927436\\
-20	0.292800699399938\\
-18	0.41953399469547\\
-16	0.579121529765205\\
-14	0.765631701178304\\
-12	0.966339686788178\\
-10	1.16454439444863\\
-8	1.34437377288987\\
-6	1.49507583337359\\
-4	1.61277037096979\\
-2	1.69941815847215\\
0	1.76030936066298\\
2	1.80164267036896\\
4	1.82901766193376\\
6	1.84684509324537\\
8	1.85832522080679\\
10	1.86566385960341\\
12	1.87033287139577\\
14	1.87329439798863\\
16	1.87516924694828\\
18	1.87635469826765\\
20	1.87710366658172\\
};
\addlegendentry{\scriptsize One-bit ADCs/DACs (Th)}

\addplot [color=blue, draw=none, mark size=2.0pt, mark=o, only marks, mark options={solid, blue}]
  table[row sep=crcr]{%
-30	0.0352645017441637\\
-28	0.0551358488708129\\
-26	0.0855990804765262\\
-24	0.131425251984132\\
-22	0.198516456010563\\
-20	0.293227893285998\\
-18	0.420587772790157\\
-16	0.580782667512879\\
-14	0.768318299715691\\
-12	0.970388417834114\\
-10	1.17074239610522\\
-8	1.35221276312203\\
-6	1.50449466105201\\
-4	1.62416159093137\\
-2	1.7116728409071\\
0	1.77313842075261\\
2	1.81438049880732\\
4	1.84197709338489\\
6	1.86047513048962\\
8	1.87222857904211\\
10	1.8800832545219\\
12	1.88454510978422\\
14	1.88747609077783\\
16	1.88887109078445\\
18	1.89016698035002\\
20	1.89108747937568\\
};
\addlegendentry{\scriptsize One-bit ADCs/DACs (MC)}

\addplot [color=mycolor2, line width=1.0pt]
  table[row sep=crcr]{%
-30	0.0906040976833155\\
-28	0.139848139941451\\
-26	0.212964903002752\\
-24	0.318201609574236\\
-22	0.463470291536921\\
-20	0.653695664809291\\
-18	0.887802687322115\\
-16	1.1569980215583\\
-14	1.44568935429619\\
-12	1.73477970173096\\
-10	2.00576528558844\\
-8	2.2441579371709\\
-6	2.4415081672427\\
-4	2.5958087058113\\
-2	2.71038270690073\\
0	2.79179259335206\\
2	2.84763275081474\\
4	2.88492894584248\\
6	2.90936938256498\\
8	2.92517672897568\\
10	2.93531120449763\\
12	2.94177143436238\\
14	2.9458742390867\\
16	2.94847367611567\\
18	2.95011811785726\\
20	2.95115741337688\\
};
\addlegendentry{\scriptsize FR ADCs/DACs (Th)}

\addplot [color=mycolor2, draw=none, mark size=2.0pt, mark=o, only marks, mark options={solid, mycolor2}]
  table[row sep=crcr]{%
-30	0.0906384681556509\\
-28	0.139900668777998\\
-26	0.213145925762648\\
-24	0.318681292053196\\
-22	0.464401573813924\\
-20	0.655511201560597\\
-18	0.891284362984778\\
-16	1.16218849357007\\
-14	1.45330711450214\\
-12	1.74521606414502\\
-10	2.01944054241573\\
-8	2.260495968313\\
-6	2.4601400413424\\
-4	2.61732382810787\\
-2	2.73381390708178\\
0	2.81504761777585\\
2	2.87194566038674\\
4	2.90970525684322\\
6	2.93537170967994\\
8	2.95101083669436\\
10	2.96166379537778\\
12	2.96779495558877\\
14	2.97231408411217\\
16	2.97434821227611\\
18	2.9757679216272\\
20	2.97711222658066\\
};
\addlegendentry{\scriptsize FR ADCs/DACs (MC)}

\addplot [color=blue, line width=1.0pt, dashdotted]
  table[row sep=crcr]{%
-30	0.0423986744452239\\
-28	0.0663365569277685\\
-26	0.103058520884926\\
-24	0.158425545125451\\
-22	0.239827451497151\\
-20	0.355355065667778\\
-18	0.511774960709262\\
-16	0.711395963019971\\
-14	0.949000792585262\\
-12	1.21083017219723\\
-10	1.47694782424023\\
-8	1.72633800238115\\
-6	1.94243356377838\\
-4	2.11660228294518\\
-2	2.24836748966588\\
0	2.34300021005422\\
2	2.40828923345327\\
4	2.45203087378096\\
6	2.48074151757623\\
8	2.49932677750588\\
10	2.51124789945867\\
12	2.51884907155592\\
14	2.52367722979371\\
16	2.52673651932512\\
18	2.52867198052464\\
20	2.52989524464263\\
};

\addplot [color=blue, dashdotted, draw=none, mark size=2.0pt, mark=o, only marks, mark options={solid, blue}]
  table[row sep=crcr]{%
-30	0.0424079725115672\\
-28	0.0663337028433817\\
-26	0.103038137814243\\
-24	0.158518226820255\\
-22	0.239924852237091\\
-20	0.355733183985511\\
-18	0.512309353769393\\
-16	0.712378898218992\\
-14	0.950606021396366\\
-12	1.21377716476352\\
-10	1.48127273567579\\
-8	1.73211536994247\\
-6	1.94912922205702\\
-4	2.12427470705924\\
-2	2.25807666003685\\
0	2.35287391059613\\
2	2.41826829844351\\
4	2.46278505803371\\
6	2.49109900596642\\
8	2.51125218174626\\
10	2.52344881050685\\
12	2.53015400520818\\
14	2.53471382089839\\
16	2.53885481139451\\
18	2.54144070879864\\
20	2.54105852052647\\
};

\addplot [color=mycolor2, dashdotted, line width=1.0pt]
  table[row sep=crcr]{%
-30	0.108901276599805\\
-28	0.168179125880781\\
-26	0.256340339000994\\
-24	0.383604806072297\\
-22	0.560191511384233\\
-20	0.79348045036086\\
-18	1.0847745309888\\
-16	1.42734776296262\\
-14	1.80702308815958\\
-12	2.17987295124195\\
-10	2.51927511102612\\
-8	2.84992834419441\\
-6	3.14358370605289\\
-4	3.3954347929807\\
-2	3.58753333488752\\
0	3.72025923468143\\
2	3.80520371993012\\
4	3.86951047721484\\
6	3.92240613335416\\
8	3.96601771760086\\
10	3.99450817110437\\
12	4.01406769268438\\
14	4.0310940349285\\
16	4.04487952554618\\
18	4.05591430927152\\
20	4.05894546562534\\
};

\addplot [color=mycolor2, draw=none, mark size=2.0pt, mark=o, only marks, mark options={solid, mycolor2}]
  table[row sep=crcr]{%
-30	0.108898275709861\\
-28	0.168240004588466\\
-26	0.256376799389347\\
-24	0.384089051184932\\
-22	0.560791157530649\\
-20	0.794724717362333\\
-18	1.08668641127387\\
-16	1.43023911663062\\
-14	1.81160496495538\\
-12	2.21108049661228\\
-10	2.52240011102612\\
-8	2.85117834419441\\
-6	3.14545870605289\\
-4	3.4035597929807\\
-2	3.60003333488752\\
0	3.74150923468143\\
2	3.8343645371993\\
4	3.89451047721484\\
6	3.94974061333542\\
8	3.99260177176009\\
10	4.01950817110437\\
12	4.03906769268439\\
14	4.05796903492851\\
16	4.06919879525546\\
18	4.08091430927152\\
20	4.08394546562534\\
};
\node at (axis cs: -29,3.8) [anchor = west] { Solid lines: $L=4$};
\node at (axis cs: -29,4.1) [anchor = west] { Dashdotted lines: $L=2$};

\end{axis}
\end{tikzpicture}%
\caption{Sum average rate versus $P_t$ for $M=128$ and $K=8$. }
\label{Fig2}
\end{minipage}
\hspace{.1cm}
\begin{minipage}[b]{0.32\linewidth}
\centering
\tikzset{every picture/.style={scale=.95}, every node/.style={scale=.8}}
%
%
\definecolor{mycolor1}{rgb}{0.60000,0.20000,0.00000}%
\begin{tikzpicture}

\begin{axis}[%
width=.85\textwidth,
height=.7\textwidth,
scale only axis,
xmin=0,
xmax=800,
xlabel style={font=\color{white!15!black}},
xlabel={$M$},
ymin=1,
ymax=4.8,
ylabel style={font=\color{white!15!black}},
ylabel={Sum average rate per user (bps/Hz)},
axis background/.style={fill=white},
xmajorgrids,
ymajorgrids,
legend style={at={(axis cs: 800,1)},anchor=south east,legend cell align=left,align=left,draw=white!15!black, /tikz/column 2/.style={
                column sep=5pt,
            }},]
\addplot [color=blue,  line width=1.0pt, mark size=2.0pt, mark=x, mark options={solid, blue}]
  table[row sep=crcr]{%
50	1.04369424613222\\
100	1.64860587610494\\
200	2.24802190720718\\
300	2.5762902781572\\
400	2.79475656513921\\
600	3.07943949171245\\
800	3.26390889046326\\
};
\addlegendentry{\scriptsize One-bit ADCs/DACs (Th)}

\addplot [color=blue, draw=none, mark size=2.0pt, mark=o, only marks, mark options={solid, blue}]
  table[row sep=crcr]{%
50	1.05999026589161\\
100	1.66648806439388\\
200	2.25918567350385\\
300	2.58919648770046\\
400	2.80326129586142\\
600	3.0846893435776\\
800	3.26933898273422\\
};
\addlegendentry{\scriptsize One-bit ADCs/DACs (MC)}

\addplot [color=red,  line width=1.0pt, mark size=2.0pt,  mark=+, mark options={solid, red}]
  table[row sep=crcr]{%
50	2.15370529034172\\
100	2.74923974031795\\
200	3.23883030035118\\
300	3.47945625376747\\
400	3.63068365448052\\
600	3.81714868945491\\
800	3.93098471660314\\
};
\addlegendentry{\scriptsize FR  ADCs/DACs (Th)}

\addplot [color=red, draw=none, mark size=2.0pt, mark=o, only marks,mark options={solid, red}]
  table[row sep=crcr]{%
50	2.19977502590063\\
100	2.78053612024587\\
200	3.25662141640835\\
300	3.4977633753808\\
400	3.6447351460794\\
600	3.82671989213574\\
800	3.93962119487599\\
};
\addlegendentry{\scriptsize FR  ADCs/DACs (MC)}

\addplot [color=mycolor1,  line width=1.0pt, mark size=2.0pt, mark=diamond, mark options={solid, mycolor1}]
  table[row sep=crcr]{%
50	4.46712632793422\\
100	4.46712632793422\\
200	4.46712632793422\\
300	4.46712632793422\\
400	4.46712632793422\\
600	4.46712632793422\\
800	4.46712632793422\\
};
\addlegendentry{\scriptsize  $M\rightarrow \infty$ (Corollary \ref{Cor_ZFlimit})}

\end{axis}
%
%
\end{tikzpicture}%
\caption{Sum average rate versus $M$ for $L=4$, $K=8$, and $P_t=10\rm{dB}$.}
\label{Fig3b}
\end{minipage}
\hspace{.1cm}
\begin{minipage}[b]{0.32\linewidth}
\centering
\tikzset{every picture/.style={scale=.95}, every node/.style={scale=.8}}
%
%
\definecolor{mycolor1}{rgb}{0.85000,0.32500,0.09800}%
\definecolor{mycolor2}{rgb}{0.00000,0.49804,0.00000}%
\definecolor{mycolor3}{rgb}{0.74902,0.00000,0.74902}%
\definecolor{mycolor4}{rgb}{0.46667,0.67451,0.18824}%
\definecolor{mycolor5}{rgb}{0.60000,0.20000,0.00000}%
\begin{tikzpicture}

\begin{axis}[%
width=.85\textwidth,
height=.7\textwidth,
scale only axis,
xmin=-30,
xmax=20,
xlabel style={font=\color{white!15!black}},
xlabel={Average transmit power $P_t$ ($\rm{dB}$)},
ymin=1,
ymax=4.7,
ylabel style={font=\color{white!15!black}},
ylabel={$\kappa$},
axis background/.style={fill=white},
xmajorgrids,
ymajorgrids,
legend style={at={(axis cs: -30,4.7)},anchor=north west,legend cell align=left,align=left,draw=white!15!black, /tikz/column 2/.style={
                column sep=5pt,
            }},]

\addplot [color=mycolor1, line width=0.8pt, mark size=1.2pt,mark=diamond, mark options={solid, mycolor1}]
  table[row sep=crcr]{%
-30	2.39\\
-29	2.4\\
-28	2.41\\
-27	2.42\\
-26	2.42\\
-25	2.43\\
-24	2.44\\
-23	2.45\\
-22	2.46\\
-21	2.47\\
-20	2.48\\
-19	2.49\\
-18	2.51\\
-17	2.53\\
-16	2.56\\
-15	2.59\\
-14	2.62\\
-13	2.66\\
-12	2.71\\
-11	2.76\\
-10	2.82\\
-9	2.89\\
-8	2.96\\
-7	3.03\\
-6	3.11\\
-5	3.19\\
-4	3.26\\
-3	3.33\\
-2	3.4\\
-1	3.46\\
0	3.51\\
1	3.56\\
2	3.6\\
3	3.64\\
4	3.67\\
5	3.69\\
6	3.71\\
7	3.73\\
8	3.74\\
9	3.75\\
10	3.76\\
11	3.77\\
12	3.77\\
13	3.78\\
14	3.78\\
15	3.78\\
16	3.78\\
17	3.79\\
18	3.79\\
19	3.79\\
20	3.79\\
};
\addlegendentry{\tiny $M^{\rm conv}=10^2$}

\addplot [color=black, line width=0.8pt, mark size=1.2pt, mark=x, mark options={solid, black}]
  table[row sep=crcr]{%
-30	2.47\\
-29	2.47\\
-28	2.47\\
-27	2.48\\
-26	2.48\\
-25	2.48\\
-24	2.49\\
-23	2.49\\
-22	2.5\\
-21	2.51\\
-20	2.52\\
-19	2.54\\
-18	2.55\\
-17	2.57\\
-16	2.6\\
-15	2.63\\
-14	2.66\\
-13	2.7\\
-12	2.75\\
-11	2.8\\
-10	2.86\\
-9	2.93\\
-8	3.01\\
-7	3.09\\
-6	3.17\\
-5	3.25\\
-4	3.33\\
-3	3.41\\
-2	3.48\\
-1	3.55\\
0	3.61\\
1	3.66\\
2	3.7\\
3	3.74\\
4	3.78\\
5	3.8\\
6	3.82\\
7	3.84\\
8	3.86\\
9	3.87\\
10	3.88\\
11	3.89\\
12	3.89\\
13	3.9\\
14	3.9\\
15	3.91\\
16	3.91\\
17	3.91\\
18	3.91\\
19	3.91\\
20	3.91\\
};
\addlegendentry{\tiny $M^{\rm conv}=10^3$}

%
%

\addplot [color=mycolor3, line width=0.8pt, mark size=1pt, mark=o, mark options={solid, mycolor3}]
  table[row sep=crcr]{%
-30	2.47\\
-29	2.47\\
-28	2.47\\
-27	2.48\\
-26	2.48\\
-25	2.48\\
-24	2.48\\
-23	2.49\\
-22	2.49\\
-21	2.5\\
-20	2.51\\
-19	2.52\\
-18	2.53\\
-17	2.55\\
-16	2.56\\
-15	2.59\\
-14	2.61\\
-13	2.65\\
-12	2.68\\
-11	2.73\\
-10	2.77\\
-9	2.83\\
-8	2.88\\
-7	2.94\\
-6	3.01\\
-5	3.07\\
-4	3.13\\
-3	3.19\\
-2	3.24\\
-1	3.29\\
0	3.33\\
1	3.37\\
2	3.4\\
3	3.42\\
4	3.45\\
5	3.47\\
6	3.48\\
7	3.49\\
8	3.5\\
9	3.51\\
10	3.52\\
11	3.52\\
12	3.53\\
13	3.53\\
14	3.53\\
15	3.54\\
16	3.54\\
17	3.54\\
18	3.54\\
19	3.54\\
20	3.54\\
};
\addlegendentry{\tiny $M^{\rm conv}=5\cdot 10^4$}

\addplot [color=blue, line width=0.8pt, mark size=1.2pt, mark=+, mark options={solid, blue}]
  table[row sep=crcr]{%
-30	2.47\\
-29	2.47\\
-28	2.47\\
-27	2.47\\
-26	2.47\\
-25	2.48\\
-24	2.48\\
-23	2.48\\
-22	2.49\\
-21	2.49\\
-20	2.5\\
-19	2.5\\
-18	2.51\\
-17	2.53\\
-16	2.54\\
-15	2.56\\
-14	2.58\\
-13	2.6\\
-12	2.63\\
-11	2.66\\
-10	2.69\\
-9	2.73\\
-8	2.77\\
-7	2.81\\
-6	2.86\\
-5	2.9\\
-4	2.94\\
-3	2.98\\
-2	3.01\\
-1	3.04\\
0	3.07\\
1	3.09\\
2	3.11\\
3	3.13\\
4	3.15\\
5	3.16\\
6	3.17\\
7	3.17\\
8	3.18\\
9	3.19\\
10	3.19\\
11	3.19\\
12	3.2\\
13	3.2\\
14	3.2\\
15	3.2\\
16	3.2\\
17	3.2\\
18	3.2\\
19	3.2\\
20	3.2\\
};
\addlegendentry{\tiny $M^{\rm conv}=10^5$}

\addplot [color=mycolor5, line width=0.8pt, mark size=1.2pt, mark=star, mark options={solid, mycolor5}]
  table[row sep=crcr]{%
-30	2.46\\
-29	2.46\\
-28	2.46\\
-27	2.46\\
-26	2.45\\
-25	2.45\\
-24	2.45\\
-23	2.44\\
-22	2.43\\
-21	2.43\\
-20	2.42\\
-19	2.41\\
-18	2.39\\
-17	2.38\\
-16	2.36\\
-15	2.34\\
-14	2.31\\
-13	2.28\\
-12	2.25\\
-11	2.22\\
-10	2.18\\
-9	2.14\\
-8	2.11\\
-7	2.07\\
-6	2.04\\
-5	2.01\\
-4	1.98\\
-3	1.95\\
-2	1.93\\
-1	1.91\\
0	1.89\\
1	1.88\\
2	1.87\\
3	1.86\\
4	1.85\\
5	1.84\\
6	1.84\\
7	1.84\\
8	1.83\\
9	1.83\\
10	1.83\\
11	1.83\\
12	1.82\\
13	1.82\\
14	1.82\\
15	1.82\\
16	1.82\\
17	1.82\\
18	1.82\\
19	1.82\\
20	1.82\\
};
\addlegendentry{\tiny $M^{\rm conv}=5\cdot 10^5$}

\addplot [color=mycolor4, line width=0.8pt, mark size=1.2pt, mark=triangle, mark options={solid, rotate=270, mycolor4}]
  table[row sep=crcr]{%
-30	2.45\\
-29	2.45\\
-28	2.44\\
-27	2.44\\
-26	2.43\\
-25	2.42\\
-24	2.41\\
-23	2.39\\
-22	2.37\\
-21	2.35\\
-20	2.33\\
-19	2.3\\
-18	2.26\\
-17	2.22\\
-16	2.17\\
-15	2.11\\
-14	2.05\\
-13	1.98\\
-12	1.91\\
-11	1.84\\
-10	1.76\\
-9	1.69\\
-8	1.62\\
-7	1.56\\
-6	1.5\\
-5	1.45\\
-4	1.4\\
-3	1.36\\
-2	1.33\\
-1	1.3\\
0	1.28\\
1	1.26\\
2	1.25\\
3	1.23\\
4	1.22\\
5	1.21\\
6	1.21\\
7	1.2\\
8	1.2\\
9	1.2\\
10	1.19\\
11	1.19\\
12	1.19\\
13	1.19\\
14	1.19\\
15	1.19\\
16	1.18\\
17	1.18\\
18	1.18\\
19	1.18\\
20	1.18\\
};
\addlegendentry{\tiny $M^{\rm conv}=10^6$}

\end{axis}
\end{tikzpicture}%
\caption{Ratio  $\kappa=\frac{M^{\rm one}}{M^{\rm conv}}$ versus $P_t$ for $L=4$ and $K=8$.}
\label{Fig5}
\end{minipage}
\end{figure*}

Next in Fig. \ref{Fig3b}, we plot the sum average rate per user against $M$ for both one-bit and conventional massive MIMO systems. For $M=800$, one-bit quantized and conventional ZF precoding are seen to achieve $73\%$ and $88\%$ of the asymptotic sum average rate outlined in Corollary \ref{Cor_ZFlimit}. The performance gap between the two settings decreases with $M$ implying that the impact of quantization becomes increasingly small as $M\rightarrow \infty$. Further to achieve the rate of $3\rm{bps/Hz}$, $M^{\rm one} = 540$  antennas should be employed at each BS of a one-bit  system, compared with $M^{\rm conv}=150$ antennas at each BS of a conventional system, implying that $\kappa=\frac{M^{\rm one}}{M^{\rm conv}}=3.6$. 




The relationship between the number of antennas  $M^{\rm one}$ needed by the one-bit MIMO system to perform as well as the conventional MIMO system with  $M^{\rm conv}$ antennas is further illustrated in Fig. \ref{Fig5}. We numerically find and plot the ratio  $\kappa=\frac{M^{\rm one}}{M^{\rm conv}}$ needed to achieve $|R_{\rm sum}^{\rm one}-R_{\rm sum}^{\rm conv}|\leq \epsilon$ for $\epsilon=10^{-3}$ and different values of $M^{\rm conv}$. The ratio is  around $2.5$  at low $P_t$ (or SNR$=\frac{P_t}{\sigma^2}$) values in accordance with Corollary \ref{Corspec1}, while  it increases to $3.79$ for $M^{\rm conv}=100$ as  $P_t$ increases to $20\rm {dB}$, due to reasons discussed in  Sec. IV-C. The promising observation is that even at moderate to high SNR values as $M^{\rm conv}$ increases, $\kappa$ increases  at a slower rate and eventually starts to decrease and approach one,  because  the effect of quantization  decreases with $M$ as discussed in Remark 1. 





\begin{figure}[!t]
\centering
\tikzset{every picture/.style={scale=.95}, every node/.style={scale=.8}}
%
%
\definecolor{mycolor1}{rgb}{0.00000,0.44700,0.74100}%
\definecolor{mycolor2}{rgb}{0.85000,0.32500,0.09800}%
\definecolor{mycolor3}{rgb}{0.92900,0.69400,0.12500}%
\definecolor{mycolor4}{rgb}{0.49400,0.18400,0.55600}%
\begin{tikzpicture}

\begin{axis}[%
width=.35\textwidth,
height=.17\textwidth,
scale only axis,
xmin=0,
xmax=400,
xlabel style={font=\color{white!15!black}},
xlabel={$f_s$ (MHz)},
ymin=1,
ymax=5.5,
ylabel style={font=\color{white!15!black}},
ylabel={EE (bps/Hz/Joule)},
axis background/.style={fill=white},
xmajorgrids,
ymajorgrids,
legend style={at={(axis cs: 400,5.5)},anchor=north east,legend cell align=left,align=left,draw=white!15!black, /tikz/column 2/.style={
                column sep=5pt,
            }},]

\addplot [color=mycolor3, line width=1.0pt, mark=star, mark options={solid, mycolor3}]
  table[row sep=crcr]{%
20	3.42438242079769\\
40	3.42198455070058\\
60	3.41959003639551\\
80	3.41719887084282\\
100	3.4148110470225\\
120	3.41242655793419\\
140	3.41004539659705\\
160	3.4076675560497\\
180	3.40529302935018\\
200	3.40292180957586\\
220	3.40055388982338\\
240	3.39818926320857\\
260	3.39582792286641\\
280	3.39346986195095\\
300	3.39111507363523\\
320	3.38876355111123\\
340	3.38641528758979\\
360	3.38407027630059\\
380	3.38172851049201\\
400	3.37938998343113\\
};
\addlegendentry{ One-bit ADCs/DACs ($M^{\rm one}=486$)}

\addplot [color=mycolor4, line width=1.0pt, mark=triangle, mark options={solid, rotate=270, mycolor4}]
  table[row sep=crcr]{%
20	5.4698267605303\\
40	4.75297144105054\\
60	4.20224120846569\\
80	3.76588535353133\\
100	3.41162605905826\\
120	3.11828678484776\\
140	2.87139756997879\\
160	2.66073486119673\\
180	2.47887027630421\\
200	2.32027645316719\\
220	2.18075549683623\\
240	2.05706193928238\\
260	1.94664712716331\\
280	1.84748173742204\\
300	1.75792991142747\\
320	1.67665830286478\\
340	1.60256923986501\\
360	1.53475085801915\\
380	1.4724393811053\\
400	1.41499023181669\\
};
\addlegendentry{  FR ADCs/DACs ($M^{\rm conv}=128$)}

\end{axis}
\end{tikzpicture}%
\caption{EE versus the sampling frequency of ADCs/DACs for $L=4$, $K=8$ and $P_t=10\rm{dB}$. }
\label{Fig5b}
\end{figure}
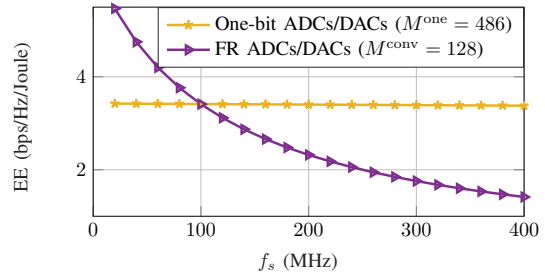

Next we study whether we gain in terms of energy efficiency (EE) when we use one-bit ADCs and DACs instead of FR ADCs and DACs. EE is defined in the downlink as $\rm{EE}=\frac{R_{\rm sum}}{P_{\rm tot}}$, where $P_{\rm tot}=\frac{1}{\zeta}P_t+M(2P_{\rm DAC}+P_{\rm RF})$,  $\zeta $ is the power amplifier efficiency, $P_{\rm DAC}$ is the power consumption of each DAC, and $P_{RF}$ is the power consumption per RF chain given in \cite{addref3}. $P_{\rm DAC}$ scales linearly with the sampling frequency $f_s$, and exponentially with the number of bits $b$ and is given as $P_{\rm DAC}=cf_s 2^b$, where $c=494$fJ/step/Hz. To compute the EE of  conventional massive MIMO cellular system, we consider $M^{\rm conv}=128$ antennas at each BS, and assume that each DAC has a resolution of $b=10$ bits to achieve nearly FR. For the one-bit massive MIMO cellular system, we find that $M^{\rm one}=486$ antennas are needed  to  achieve the same sum average rate  as the conventional  system. Using this value, we compute  $R_{\rm sum}^{\rm one}$, $P_{\rm tot}$ with $b=1$, and consequently the EE.  The results are plotted against $f_s$ in Fig. \ref{Fig5b}. The EE achieved by the one-bit MIMO system exceeds that achieved by the conventional MIMO system for $f_s>100$MHz, while achieving the same sum average rate. The decrease in the EE of the conventional system with $f_s$ is significant because  the power consumption of FR  ADCs and DACs is quite dominant. This is a very promising result especially for mmWave communication systems, that utilize larger bandwidths and higher sampling rates. Thus one-bit massive MIMO is an energy-efficient solution even under linear ZF precoding and imperfect CSI for mmWave systems.




\vspace{-.14in}

\section{Conclusion}
\label{Sec:Con}
\vspace{-.07in}

This work studied a multi-cell massive MIMO system employing one-bit ADCs and DACs under   ZF precoding  and imperfect CSI. We  derived  closed-form expressions of the MMSE channel estimates at each BS and the ergodic achievable downlink rates at the users, utilizing the Bussgang decompositions of the quantized received  training and  transmit signals  respectively. We then studied the ratio of the number of antennas at each BS in the one-bit cellular system to that at each BS in the conventional system required for both systems to achieve the same sum average rate.  The ratio turned out to be  $2.5$  at low SNR,   while it was seen to decrease to one for any given SNR as we consider larger antenna arrays.  We also observed  one-bit  MIMO to be more energy efficient than conventional MIMO at higher bandwidths. 

\vspace{-.05in}
\vspace{-.1in}

\bibliographystyle{IEEEtran}
\bibliography{bib}

\begin{thebibliography}{10}
\providecommand{\url}[1]{#1}
\csname url@samestyle\endcsname
\providecommand{\newblock}{\relax}
\providecommand{\bibinfo}[2]{#2}
\providecommand{\BIBentrySTDinterwordspacing}{\spaceskip=0pt\relax}
\providecommand{\BIBentryALTinterwordstretchfactor}{4}
\providecommand{\BIBentryALTinterwordspacing}{\spaceskip=\fontdimen2\font plus
\BIBentryALTinterwordstretchfactor\fontdimen3\font minus
  \fontdimen4\font\relax}
\providecommand{\BIBforeignlanguage}[2]{{%
\expandafter\ifx\csname l@#1\endcsname\relax
\typeout{** WARNING: IEEEtran.bst: No hyphenation pattern has been}%
\typeout{** loaded for the language `#1'. Using the pattern for}%
\typeout{** the default language instead.}%
\else
\language=\csname l@#1\endcsname
\fi
#2}}
\providecommand{\BIBdecl}{\relax}
\BIBdecl

\bibitem{massiveMIMOO}
J.~Hoydis, S.~ten Brink, and M.~Debbah, ``Massive{ MIMO in the UL/DL} of
  cellular networks: How many antennas do we need?'' \emph{IEEE J. Sel. Areas
  Commun.}, vol.~31, no.~2, pp. 160--171, 2013.

\bibitem{ADC1}
R.~Walden, ``Analog-to-digital converter survey and analysis,'' \emph{IEEE J.
  Sel. Areas Commun.}, vol.~17, no.~4, pp. 539--550, 1999.

\bibitem{lit12}
A.~K. Saxena, I.~Fijalkow, and A.~L. Swindlehurst, ``Analysis of one-bit
  quantized precoding for the multiuser massive {MIMO} downlink,'' \emph{IEEE
  Trans. Signal Process.}, vol.~65, no.~17, pp. 4624--4634, 2017.

\bibitem{lit9}
Y.~Li \emph{et~al.}, ``Channel estimation and performance analysis of one-bit
  massive {MIMO} systems,'' \emph{IEEE Trans. Signal Process.}, vol.~65,
  no.~15, pp. 4075--4089, 2017.

\bibitem{addref}
S.~Jacobsson, G.~Durisi, M.~Coldrey, T.~Goldstein, and C.~Studer, ``Quantized
  precoding for massive {MU-MIMO},'' \emph{IEEE Trans. Commun.}, vol.~65,
  no.~11, pp. 4670--4684, 2017.

\bibitem{lit10}
Y.~Li, C.~Tao, A.~Lee~Swindlehurst, A.~Mezghani, and L.~Liu, ``Downlink
  achievable rate analysis in massive mimo systems with one-bit {DACs},''
  \emph{IEEE Commun. Lett.}, vol.~21, no.~7, pp. 1669--1672, 2017.

\bibitem{lit11}
Y.~Zhang \emph{et~al.}, ``Rate analysis of cell-free massive {MIMO} with
  one-bit {ADCs and DACs},'' in \emph{IEEE International Symposium on Personal,
  Indoor and Mobile Radio Communications (PIMRC)}, 2019, pp. 1--6.

\bibitem{addref3}
E.~Balti and B.~L. Evans, ``A unified framework for full-duplex massive {MIMO}
  cellular networks with low resolution data converters,'' \emph{IEEE Open J.
  Commun. Soc.}, vol.~4, pp. 1--28, 2023.

\bibitem{Bussgang}
J.~J. Bussgang, ``Crosscorrelation functions of amplitude-distorted gaussian
  signals,'' Res. Lab. Electron., Massachusetts Inst. Technol., Cambridge, MA,
  USA, Tech. Rep. 216, 1952.

\bibitem{Bussgang1}
A.~Papoulis and S.~U. Pillai, \emph{Probability, Random Variables, and
  Stochastic Processes}.\hskip 1em plus 0.5em minus 0.4em\relax New York, NY,
  USA: McGraw-Hill, 2002.

\bibitem{CLT}
H.~Cramer, \emph{Random Variables and Probability Distributions, vol.
  36}.\hskip 1em plus 0.5em minus 0.4em\relax Cambridge, U.K: Cambridge Univ.
  Press, 2004.

\bibitem{AR}
J.~Jose, A.~Ashikhmin, T.~L. Marzetta, and S.~Vishwanath, ``Pilot contamination
  and precoding in multi-cell {TDD} systems,'' \emph{IEEE Trans. Wirel.
  Commun.}, vol.~10, no.~8, pp. 2640--2651, 2011.

\end{thebibliography}
\vspace{-.1in}
\end{document}